\documentclass{article}

\makeatletter
    \let\@fnsymbol\@alph
\makeatother
\usepackage{caption}
\usepackage{arxiv}

\usepackage[utf8]{inputenc} 
\usepackage[T1]{fontenc}    
\usepackage{hyperref}       
\usepackage{url}            
\usepackage{booktabs}       
\usepackage{amsfonts}       
\usepackage{nicefrac}       
\usepackage{microtype}      
\usepackage{lipsum}		
\usepackage{graphicx}
\usepackage{natbib}
\usepackage{doi}

\usepackage{multirow}
\usepackage{colortbl}

\title{Introducing causal inference in the energy-efficient building design process}

\author{
    \href{https://orcid.org/0000-0001-8504-2303}{\hspace{1mm}Xia Chen $^*$ \thanks{Leibniz University Hannover, Germany, Technische Universität Berlin, Germany  <xia.chen@tu-berlin.de>}} \\
	\And
	\href{https://orcid.org/0000-0002-8100-9861}{\hspace{1mm}Jimmy Abualdenien\thanks{Technische Universität München, Germany}} \\
	\And
	\href{https://orcid.org/0000-0001-7921-9475}{\hspace{1mm}Manav Mahan Singh\thanks{Technische Universität München, Germany, Katholieke Universiteit Leuven, Belgium}} \\
	\And
	\href{https://orcid.org/0000-0003-2088-7254}{\hspace{1mm}André Borrmann\footnotemark[2]} \\
	\And
	\href{https://orcid.org/0000-0002-0935-4361}{\hspace{1mm}Philipp Geyer\thanks{Leibniz University Hannover, Germany, Technische Universität Berlin, Germany}} \\
}

\renewcommand{\shorttitle}{Introducing causal inference in the energy-efficient building design process}

\hypersetup{
pdftitle={Introducing causal inference in the energy-efficient building design process},
pdfsubject={},
pdfauthor={X Chen},
pdfkeywords={},
}

\begin{document}
\maketitle

\begin{abstract}
“What-if” questions are intuitively generated and commonly asked during the design process. Engineers and architects need to inherently conduct design decisions, progressing from one phase to another. They either use empirical domain experience, simulations, or data-driven methods to acquire consequential feedback. We take an example from an interdisciplinary domain of energy-efficient building design to argue that the current methods for decision support have limitations or deficiencies in four aspects: parametric independency identification, gaps in integrating knowledge-based and data-driven approaches, less explicit model interpretation, and ambiguous decision support boundaries. In this study, we first clarify the nature of dynamic experience in individuals and constant principal knowledge in design. Subsequently, we introduce causal inference into the domain. A four-step process is proposed to discover and analyze parametric dependencies in a mathematically rigorous and computationally efficient manner by identifying the causal diagram with interventions. The causal diagram provides a nexus for integrating domain knowledge with data-driven methods, providing interpretability and testability against the domain experience within the design space. Extracting causal structures from the data is close to the nature design reasoning process. As an illustration, we applied the properties of the proposed estimators through simulations. The paper concludes with a feasibility study demonstrating the proposed framework’s realization. 
\end{abstract}


\section{Introduction}
Sustainable building design is challenged by multi-objective and multi-variable design tasks  \citep{westermann2019surrogate, evins2013review}. Driven by the development of Building Information Modeling (BIM)  \citep{eastman2011bim} and parameterization  \citep{shiel2018parametric}, building designers tend to reach their decision by setting boundary conditions to run virtual optimization or building simulations for scheme validations. Significant energy savings can be achieved in buildings if they are properly designed \citep{cao2016building}, which raises the requirement for building designers to progress their work in an energy-efficient manner under an environmentally friendly and cost-effective consideration. This manner usually asks for interdisciplinary knowledge from architects and engineers, e.g., building envelope improvements, heating/cooling capacity reductions, etc. Such qualitative and quantitative knowledge about the design parameters and decisions in the early stage is considered most needed for assessing design schemes \citep{fernandez2020relationship}. Subsequently, numerous simulation and optimization tools emerge to support decision-making  \citep{ostergaard2016building}. It is well established that the digitalization progress benefits the design process \citep{farzaneh2019review}. Despite the advantages and achievements of effective decision-support systems driven by digitalization, we identified four major limitations existing in the current research:
\begin{itemize}
\item \textbf{Parametric independence}: Building design optimization and building performance simulation (BPS) tools use a set of design parameters for representing the building design; however, the dependencies among the different building components and their parameters are not carefully considered, taking into account their influence on simulation results. The well-known mantra in statistics, “correlation does not imply causation” \citep{pearl2018book, aldrich1995correlations}, remains under-appreciated and rarely discussed in the building design context \citep{westermann2019surrogate, seyedzadeh2018machine}.
\end{itemize}
\begin{itemize}
\item \textbf{Integration of knowledge-based and data-driven methods}: As for the methodology, two major BPS approaches, the first-principles model (known as rule-based or knowledge-based models) and the machine learning (ML) approach \citep{amasyali2018review}, similar to the debate of symbolism and connectionism (mathematical models) in the artificial intelligence (AI) community \citep{minsky1991logical}, exist in parallel. Research investigations for the method or process development to integrate both approaches are missing.
\item \textbf{Model interpretation}: The design process involves multi-variates (features of building composition, geometry, material, structure, etc.) interactions and correlations with specific sequential constraints. In practice, there is often limited explicit consequence explanation given for approaches (especially machine learning methods) \citep{geyer2021explainable, arrieta2020explainable} to assist informed decision-making.
\item \textbf{Ambiguous boundaries of decision support}: Throughout design phases, the attention of domain experts oscillates between understanding the problem and developing a solution. Although various guidelines exist for describing the expected deliverables at every milestone (such as the U.K’s RIBA \citep{harriss2015radical}), the required semantic and geometric information and their topological relationships are not explicitly defined \citep{abualdenien2019meta}. Hence, the scope of the potential design space in each phase is ambiguous, leading to a wide spectrum of performance uncertainties.
\end{itemize}

In this context, both ambiguous boundaries of decision support and the less comprehensive machine assistance methods need to be delicately defined. The optimization analysis by manipulating parameters independently without considering the sequential hidden relationship within variables brings biased results or spurious relationships. Such results will lead to false assumptions for decision-making support, which is the gap in the above-mentioned methods for flexible alternative analysis in potential design space exploration. Essentially, the design process aims to develop interactions with users to determine the most compatible adjustment with set objective(s). The design scheme adjustment is to "control" variables, receive causal effect information, and attempt to optimize the target, e.g., building energy performance, cost, user comfort, even occupants’ habits \citep{hansen2018building, santin2009effect}, etc. The interoperability and the interchange information capability from modeling are essential for optimizing design schemes, involving “what-if” scenarios analysis, and verifying the tool’s reliability \citep{fernandez2020relationship}. In this context, it is important to separate representation for the implicit, transferable feature relationships and explicit, quantifiable change impacts individually.

In this study, we introduce causal inference into the energy-efficient building design domain, by proposing a four-step process framework of causal structure finding and causal effect estimation with a nexus for integrating domain knowledge with data-driven approaches. We provide building designers a higher dimension to inspect their own design case continuously with a simple case for framework implementation. Finally, based on this work, we establish the foundation for researchers, engineers, and designers to engage in what is now called "causal modeling" and use principles of causality to encode domain knowledge with data interpretation processes in various application domains.

The remainder of this paper is organized as follows: The expansion studies of the abovementioned defects and causality introduction are presented in section \ref{Sec2}. It is followed by the necessary method support for the proposed causal process (section \ref{Sec3}). Finally, a case study of the building design process is presented in section \ref{Sec4} to illustrate the application of the causal process with data validation. Section \ref{Sec5} \& \ref{Sec6} discuss the future potential and conclude.

\section{Method: causal inference}\label{Sec2}
Initiated by medical statistic research \citep{patil1981causal}, causal inference is now a critical research topic in many domains given observational, multivariate data to reveal an effect and the cause that’s influencing it, making interventions (“What-if..”) and counterfactual reasoning (“What if I had…”) possible \citep{pearl2018book, yao2021survey}. In more detail, a process of drawing a conclusion based on the conditions of the occurrence of an effect \citep{yao2021survey, kalisch2014causal}. Causality is commonly confused with correlation, but the former presents a different interpretation from observational data: it analyzes the asymmetric change and response between cause variables and effect. Such reasoning ability is essential for informative and sequential decision-making support in the design process. In an overview, causal research focuses on two main objectives: \textit{learning causal relations} and \textit{learning causal effects} \citep{guo2020survey} with preliminaries for two general logically equivalent frameworks: \textit{Structural Causal Model (SCM)} \citep{pearl2000models} and \textit{potential outcome framework} \citep{rubin1974estimating}. For the extension introduction of causal inference, we point to Pearl \citep{pearl2009causal}, Spirtes et al. \citep{spirtes2010introduction, spirtes2000causation}, and Peters et al. \citep{peters2017elements}.

If we inspect the design process itself, the sequential series of activities is equivalent to a process of data generation and refinement. The laws of traditional probability theory do not dictate how one property of distribution ought to change with structure-based logic \citep{judea2010introduction}. If we carefully review the way how human designers think, we find that expert design relies on the same \textbf{knowledge base} (physical principles, directionality of the causality, deterministic, universal) and variety with their personal \textbf{experience} (preferences, design variation, statistic, individual). In this process, the knowledge base is reusable and stackable (especially the sequential logic in the design process) since physical laws and foundations are less likely to be changed. Subsequently, designers make changes via experience intuitively for different cases based on objective restrictions or personal preference; whereas in first-principles modeling processes, the physical knowledge (causal relationships) encoded for each simulation remains unchanged, and requires simulations for individual cases and design phases, making it cost-inefficient in practice with full-scale experiments and extended for design alternatives. 

\begin{figure}[h]
	\centering
	\includegraphics[width=13cm]{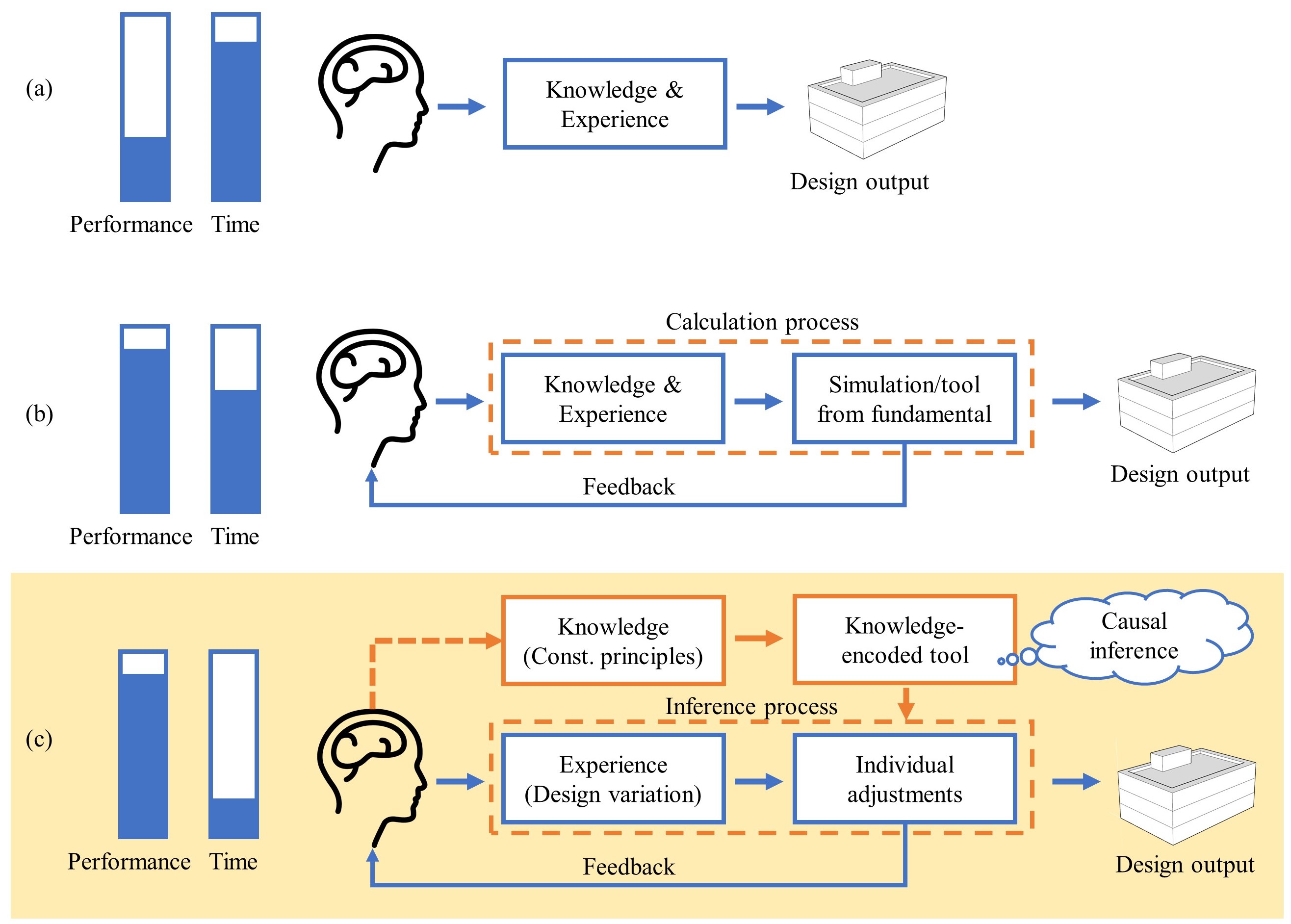}
	\caption{Different design process paradigms: (a) using only personal knowledge \& experience to design, poor performance/time trade-off because of the absence of feedback, and heavily rely on personal knowledge scope; (b) using knowledge encoded simulations/surrogate models and running them recursively, receive feedback for design exploration \& optimization, without explicit distinguish knowledge and experience, most of the decision support systems are categorized in this class, time-consuming; (c) using separated tools/adjustments to deal with knowledge \& experience, individually, loop calculation for only adjustment, knowledge could be inherited from external resources or discovered in data, enable designers to focus on the intuitive exploration based on personal experience. In this study, we emphasize inheriting knowledge of causal inferences logic (data-driven) to enhance the flexibility and efficiency of the feedback of the potential design space exploration process.}
	\label{fig:Picture1.jpg}
\end{figure}

To address this issue, we argue that it is necessary to distinguish and deal with knowledge and experience in different methods. With the conceptual framework comparison of current simulations/tools, we proposed a new paradigm of design process assistance and presented different paradigms illustratively in Figure \ref{fig:Picture1.jpg}. Essentially, it refers to a core issue of how to integrate knowledge into the modeling process so that each simulation run is based on the same cognitive foundation rather than from scratch. 

Here, we want to underline that the “knowledge-encoded tool” we defined in Figure \ref{fig:Picture1.jpg}, sub-graph (c), is different than the classical rule-based, symbol-manipulation expert system or decision support systems (DSS) \citep{liu2010integration,marcher2020decision,attia2012simulation}, which normally require a predefinition in a semantic grounding context. In fact, given the wide range of diversity and individuality in design cases, the utility of this semantic rule-based system will find it hard or computationally expensive to exhaustive all potential design spaces. The novelty we want to mention is: Instead of manually hardcoding domain rules embedded causal constraints \textbf{(first-order method for causality encoding, semantic grounding required)}, the gap we found that needs to be filled is, to make machine assistance acquire the ability to conduct cause-effect reasoning in a data-driven process. Substantially, most analysis scenarios in our domain usually come with assumption questions. In this context, a systematic generalization method to manipulate parameters freely and adapt to different cases is needed \textbf{(second-order method for causality emergence, a tool to find and encode causality directly from data, no semantic grounding required)}. To sum up, the new paradigm brings a significant advantage for the assistance: It inherits the fast and interpolable characteristics of the data-driven approach while containing the causal sequential logic for knowledge-based extraction and reuse.

In this context, causal inference reveals a possible solution of two-part objectives: \textit{causal structure finding} and \textit{causal effects estimation}. In this study, we use graph representation to clarify the estimand and causal interpretation for design adjustments, which we call causal inference.

\subsection{Causal relations: Causal structure finding}

The first pitfall in our domain research to adopt ML/AI methods is that, we usually neglect that their modeling processes often assume input parameters are independent, or even \textit{independent and identically distributed} (i.i.d.)  \citep{scholkopf2022causality} by default. That is, \textbf{the probability distribution of each value (parameter) should have no dependence on other values}; however, the design process is a series of sequential interactions between designers and the scheme (generated data) under dynamic conditions. These conditions constitute different objectives (structures, energy efficiency, emissions, etc.) within a set of shared parameters. The design process itself also contains strong nature of interventions and counterfactual assumptions embedded in the system. In other words, investigating parameters/properties without considering cause-effect logic sequence leads to unrealistic conclusions. In this case, the directionality of cause and effect should not be neglected, which is the realm of causality. In this study, we encoded the causal structure knowledge and represented it in the form of the directed acyclic graph (DAG) \citep{judea2010introduction}. DAGs are graph diagrams composed of variables (nodes) connected via unidirectional arrows (arcs) to depict hypothesized causal relationships. A directed edge $x \rightarrow y$ denotes a causal effect of $x$ (treatment) on $y$ (outcome). Intuitively, it means that $y$ is directly influenced by the status of $x$, altering $x$ by external intervention would also alter $y$.

\begin{figure}[h]
	\centering
	\includegraphics[width=15cm]{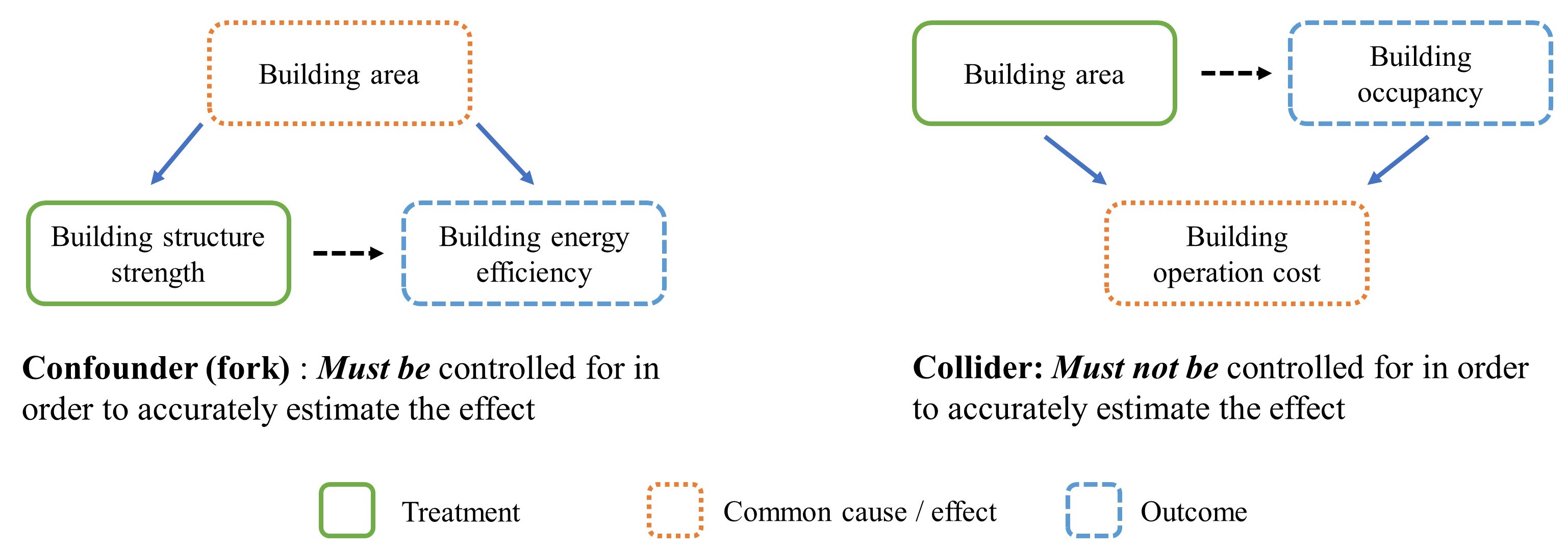}
	\caption{Causal confounder and collider examples in the building design process. Left: Building area is a common cause affecting both building structure strength and building energy efficiency, if this confounder is not controlled, it might lead to the \textbf{wrong conclusion} (“\textit{You need to adjust the building strength to change the building energy efficiency.}”); Right: Building operation cost is a common effect of both building area and building occupancy, if this collider is controlled, it leads to the spurious correlation between treatment and outcome: Building area and building occupancy are correlated (d-connected) \citep{pearl2009causal} ; Correlation should not imply the causation. }
	\label{fig:Picture2.jpg}
\end{figure}

DAGs are often defined by prior knowledge and could be incomplete \citep{guo2020survey}. They are commonly used for SCMs to express the directionality of the underlying process, and help to provide a precise graph for communicating assumptions under which the questions need to be answered to avoid spurious causal relationships. As the form shown in Figure \ref{fig:Picture2.jpg}, causal knowledge-encoded DAG helps to clarify two things in the design process:
\begin{itemize}
\item If we make “what-if” assumptions without causality inference, the confusing relationship between the measurement variables and outcome results will remain unclarified, leading  to potentially biased estimates, incorrect correlations, and finally, unrealistic potential design reasoning. 
\item The rules of causal DAGs embedded the information of conditional independence, which is identified by graph-based criteria, such as \textit{dependency separation rules (d-separation)} \citep{pearl2009causal}. Thus a DAG depicting a causal model contains genuinely more general, transferable information, which helps to rule out spurious conclusions from statistical relationships among design variables.  
\end{itemize}

The causal model with a fixed causal skeleton provides the flexibility to allow designers to base on adjusted variables and outcomes to confirm causal paths, explore different design spaces, and necessary variables that need to be conditioned. In SCMs, DAGs provide principled structural equations for identifying suitable sets of covariates for removing structural confounding bias through adjustments, e.g. \textit{back-door criterion} \citep{pearl2000models} and its extensions. By following the causal independence of the variables, biased results and unrealistic schemes are then avoided. DAG illustrations in the remaining contents of this paper are generated by DAGitty \citep{textor2016robust}.

Apart from the content of how we represent and store knowledge via DAGs with causal rules, we also need methods for knowledge extraction from data to complete the mechanism of causal structure finding. In the causal inference community, such algorithms are served as \textbf{causal discovery}: identify and return the equivalence class of the true causal structure based on observational data. They are broadly classified into three categories: \textit{approaches assuming no hidden confounders},\textit{ using a mix of observational and interventional data}, and \textit{approaches assuming hidden confounders} \citep{kalisch2014causal}. To alleviate the complexity caused by real-world data noise, potential bias, or missing information, in our case study, we use first-principles modeling simulation to clarify the estimand; hence, we present the process in the scope within approaches assuming no hidden confounders. In this context, typical causal structure finding algorithms based on observational data are categorized into:\textit{ constraint-based, score-based methods}, and \textit{hybrid methods}. Considering the building model is represented as a set of design parameters (high-dimensional sparse data), given the vast search space, greedy algorithms have been proposed \citep{kalisch2014causal}. In this study, we selected one of the typical score-based methods, \textit{Greedy-Equivalent-Search (GES)} \citep{chickering2002learning, chickering2002optimal}, to fit our design process scenario. 

\begin{figure}[h]
	\centering
	\includegraphics[width=15cm]{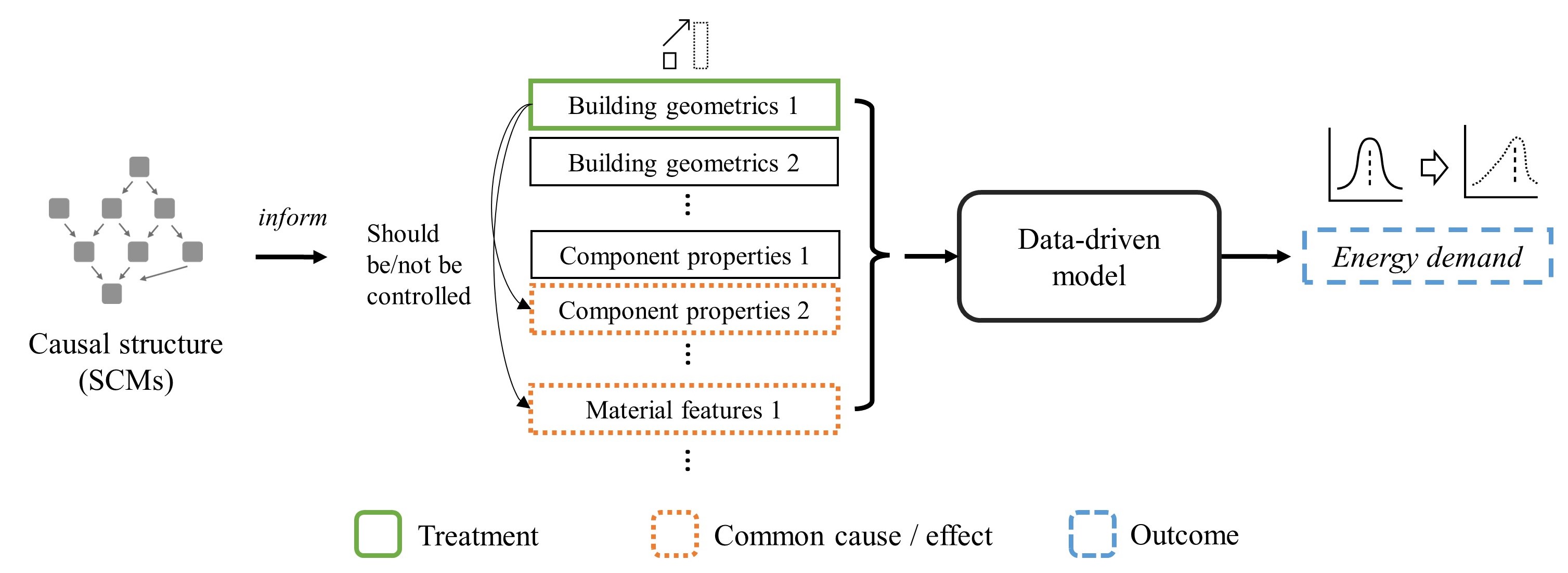}
	\caption{Causal-informed assistance case in data-driven building energy demand estimation. The data-driven model stands for the ML/AI monolithic method, in which the causal structure helps to provide insights into input dependencies and, by controlling them accordingly, to generate unbiased outcomes. }
	\label{fig:Picture3.jpg}
\end{figure}

From the practical point of view, the causal discovery with information attached causal skeleton plays a vital role in the data-driven process: It helps designers/engineers to discover domain insights among input features, overcome subjective bias, and comprehensively examine whether there are any hidden relationships neglected. Figure \ref{fig:Picture3.jpg} illustratively presents a case of how causal structure assists the data-driven model for potential outcomes: Suppose we use a monolithic black-box model (single, compact, trained model, normally refer to ML/AI methods) for potential outcomes estimation, and assume that we have no corresponding domain knowledge, SCMs could offer information regarding which input variables we should manipulate to obtain unbiased, correct consequential outcomes. 

To sum up, in the scope of causal structure finding, the SCM provides a representation of causality consisting of causal graphs (DAGs) and corresponding structural equations. A DAG contains a coherent mathematical foundation for causal structure representation followed by theories to provide a precise language for encoding the nature of causal rules. The causal structure finding algorithm converts domain knowledge to a formal model of causal assumptions and implies conditional statistical independence. Furthermore, DAGs provide a nexus for integrating data-driven approaches with domain knowledge to allow experts and building designers to examine knowledge fundamentals and change accordingly. The graph then will significantly shape the causal effect estimation.

\subsection{Causal Effect Estimation}

Based on the information of causal structure, a further important pillar of causal inference is causal effect estimation. 

As an extension of traditional statistics, the causal effect estimation investigates to what extent manipulating the value of a potential cause would influence a possible effect: A standard dataset with n instances for learning causal effects  $(X,t,y)=[(x_1,t_1,y_1),...,(x_n,t_n,y_n)]$, includes feature matrix $X$, a vector of treatments (factor that need to be controlled) $t$ and outcomes $y$. In our study, we use a more straightforward tool: potential outcome framework, a logically equivalent method to SCMs \citep{pearl2009causal}, to describe the causal effect quantification. The problem can be defined as: Given $(X,t,y)$, how the outcome $y$ is expected to change if we modify the treatment from $c$ to $t$, which denote as $\tau =E[y\mid t]- E[y\mid c]$, where $c$ to $t$ represent and the control and the treatment. 

Depending on the treatment, we care about different causal relationships and effects by conditions, sub-datasets, or individual cases. The most common treatment effect is the \textbf{Average Treatment Effect (ATE)}, which is helpful in making decisions on whether a treatment should be introduced. In terms of evaluation, regression error metrics are fit for ATE learning; however, in the design process, users care less about the ATE because each design is an individual case and potential design space is conditional on the current design scheme. In this case, \textbf{Conditional Average Treatment Effect (CATE)} is defined as to learning causal effect that consists of heterogeneous groups based on the graph information (causal relationships) presented by DAGs and graphical-based rules (e.g., do-calculus) \citep{pearl2000models}. Essentially, it allows interventions from data and ensures that the effect is correctly learned in each homogeneous group to avoid spurious or paradoxical conclusions presented in Figure \ref{fig:Picture2.jpg}; For extension understanding, we recommend \textit{Simpson's paradox} \citep{hernan2011simpson}. 

Once we properly eliminate confounding bias from data by transforming interventional distributions (block all “back-door” paths in the DAG), causal effect based on estimating the outcome model can be quantified using general regression models or ML-based algorithms. In this paper, we use simple linear regression. With \(\widehat{y}_i^t\) estimated by modeling expectation \(E[y\mid t,x_i]\), we can estimate the causal effect $\tau$ by:

\begin{equation}
    \tau =\frac{1}{n}[\sum_{i=1}^{n}( \widehat{y}_i^t-\widehat{y}_i^c)])]
\end{equation}

Thus, within the same causal structure, the causal effect (potential outcomes) under the same assumption requires only a single run of calculation. The output distribution contains the corresponding subset of $(X,t,y)$ with the information of all variate feature condition details within the set range. One huge advantage of this method lies in its low computation complexity: To conduct the same result, simulation or ML needs to run repetitively to cover possible combinations within all ranges of features. For the coding implementation, we use open-source package: \textit{DoWhy} \citep{sharma2020dowhy}.

\section{Design process with causal inference}\label{Sec3}

The design process involves a series of interactions between different entities (designer, engineer, client, etc.), which are affected by various requirements in design features, combinations of properties, concepts, needs, etc. In the building design process, designers need to make sequential decisions at different stages, containing possible building components geometry, components properties, alternative materials, and energy systems under sustainability criteria. For example, the architect and structural engineer need to collaboratively develop and exchange design information to coordinate the placement of the energy system through the required voids in structural systems. The design is typically progressed iteratively from a coarse Level of Development (LOD) to a finer one, where additional object attributes are provided or specified more accurately \citep{abualdenien2018multi}; however, in practice, the decision-making process involves invertible assumptions: Feedback is generated after the result is presented, the intervention often occurs in different scenarios: reconsideration by the designer; new conditions attached by experts’ feedback; regulations; or even client changed the idea.

\begin{figure}[h]
	\centering
	\includegraphics[width=\textwidth]{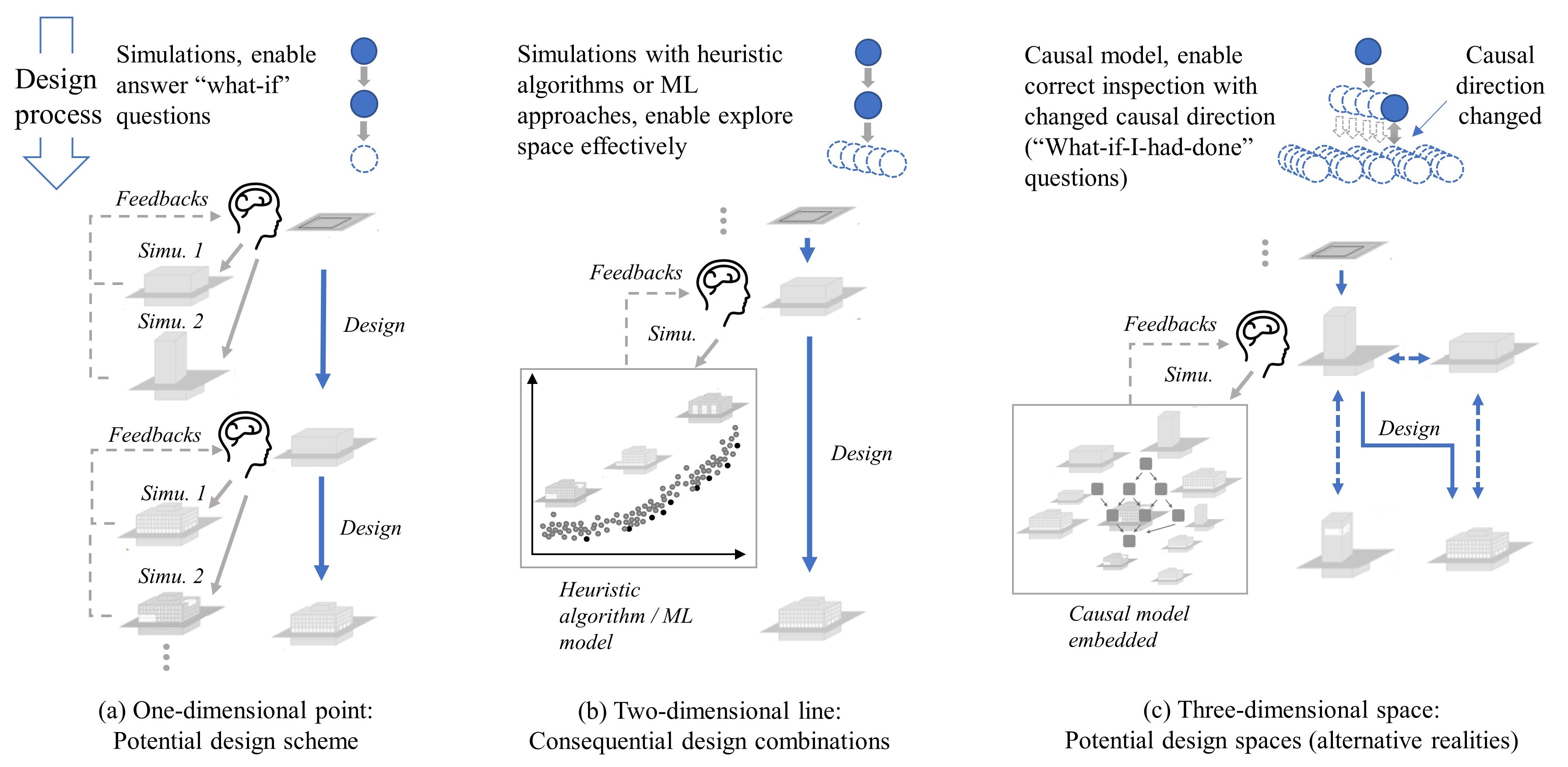}
	\caption{Design assistance dimensions: a) one-dimension: running simulations for each design scheme and receiving feedback point by point; b) two-dimension: running simulations with a heuristic algorithm or using machine learning to traverse sequential design space for optimal schemes; c) three-dimension: combining with the causal inference model would allow designers to inspect reversible, alternative design spaces for optimal schemes searching efficiently and correctly.}
	\label{fig:Picture4.jpg}
\end{figure}

In this background, BPS tools have been developed to inform the design process. Essentially, they are applied to generate consequences from assumptions to answer causal questions by setting up simulation scenarios as design assistance \citep{chen2022machine}. It is certainly beneficial for creating simulations with more details to enhance the performance accuracy \citep{de2014gap}; however, we argue that it is not only helpful but essential to allow the tool to inspect the potential design space toward a higher dimension – reversible design process assistance and alternative potential space exploration. Accordingly, we presented a summary illustration describing design assistance dimensions in Figure \ref{fig:Picture4.jpg}.

As shown in Figure \ref{fig:Picture4.jpg}, a single run of simulation via the first-principles model reveals one specific potential design scheme at the one-dimensional point (node); Optimization algorithms \citep{kheiri2018review} elevate the inspection scale to the two-dimensional consequential design space (“what-if”) by searching for combinatorial optimums to reduce the calculation complexity of repetitive simulations in parametric permutation. ML methods further accelerate the combination process by discovering and inducing relationships between variables; however, they lack corresponding mechanisms to address hidden relationships between different features and the result: multi-causes, confounders and, lurker variables \citep{pearl2018book}; Ignoring the implicit variable relationships will bring potentially biased estimates or spurious relationships, which restrain the ML methods properly applied into the third dimension that arises during the design process: the parallel potential design space (“what-if-I-had-done”). Compared to our aforementioned-invertible process with causal inference, the differences (which are usually neglected or less rigor distinguished) are:
\begin{itemize}
\item ML methods and optimization process: “I don't see the consequential result. I wonder how objective $Y$ would change if I set variable(s) $X$” – \textit{what if} – no directional change of cause and effect – based on observation (i.e. subgraphs a, b in Figure \ref{fig:Picture4.jpg})
\item Causal inference model process: "I saw the consequential result, if I take variable(s) $X$ and adjust it, how would objective $Y$ change; will there be designs (related independent combinations) that are better than the original after the adjustment?" – \textit{what if I had done} – directional change of cause and effect required – based on counterfactual intervention (i.e. subgraph c in Figure \ref{fig:Picture4.jpg})
\end{itemize}

To sum up, we argue that the causal model provides the necessary ground for developing a design space exploration mechanism. The DAG presents the cause-effect structure and enables the corresponding intervention assumptions to be made and examined during the design process. Moreover, if the assumption is within the scope of the same causal structure, the computational complexity is low to explore potential design space via a causal model inference than through exhaustive simulations. It lays the foundation of the engineering feasibility to gain real-time feedback, when the computation difficulty does not increase drastically when the number of input variables increases. The core advantage of embedding a causal model is: It allows designers to correctly inspect potential design alternatives with elevated dimensionality in a computationally attractive way. 

Meanwhile, the causal structures characterize the underlying physical composition of the building at the current design phase. As a representation of the design process at a certain stage, the causal skeleton draws a clear boundary that the building variants follow the same intellectual context to an extent, thus regulating the boundaries of design assistance. 

\begin{figure}[h]
	\centering
	\includegraphics[width=14cm]{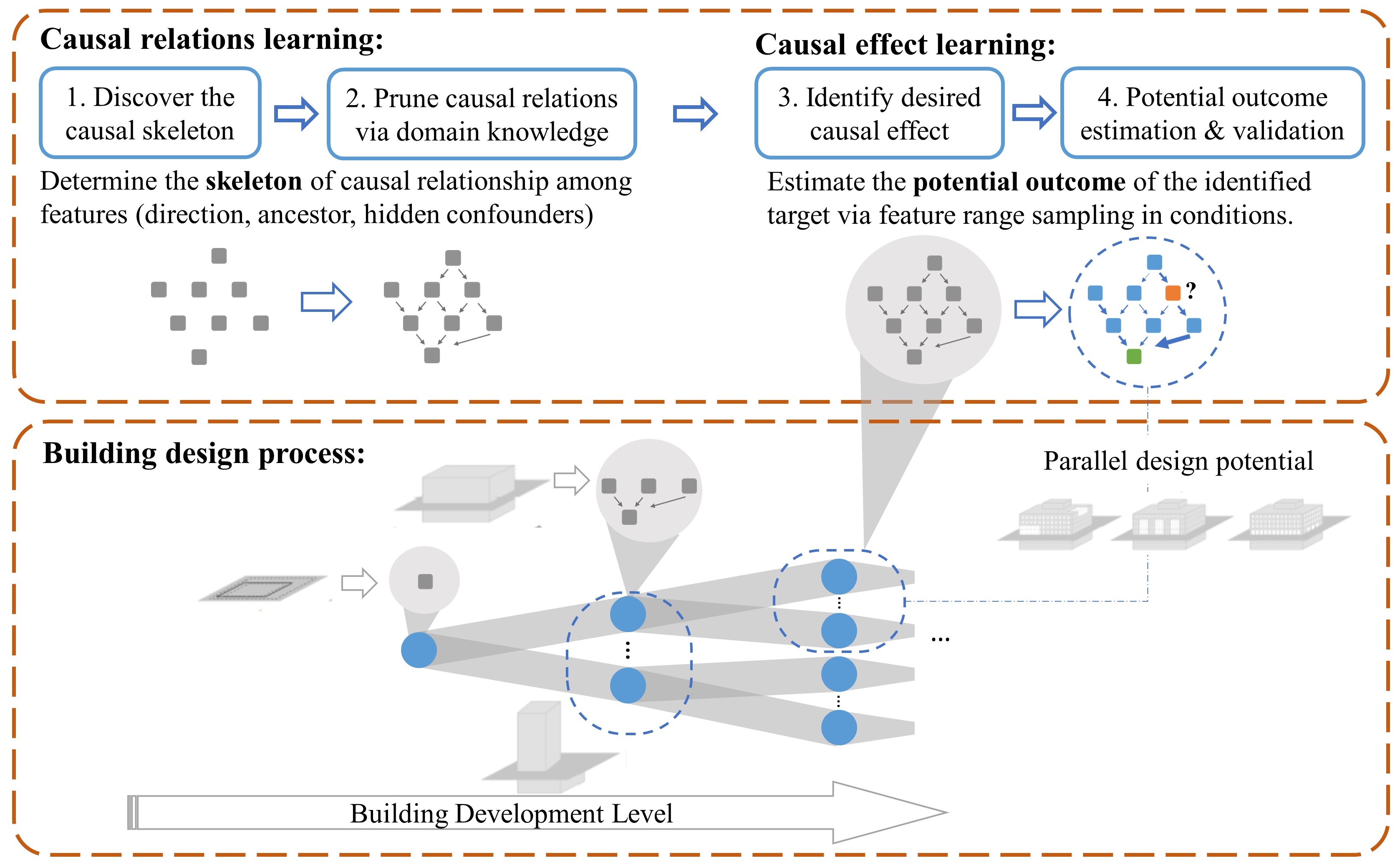}
	\caption{The four-step framework implementation of causal inference in the building design process: The first two steps of the framework aim to conduct causal relationships extraction at a certain level, and present them in a skeleton by SCMs; the process could be data-driven and/or knowledge-based. DAGs provide a medium for experts to interpret and integrate their domain knowledge. The final two steps use data-driven quantification methods (Robin causal model) to first identify the target based on conditions extract from the causal skeleton, which uses graph-based criteria. Then target causal effect is estimated with interventional outcomes in potential design space (alternative realities) by running a single calculation based on the assumption. The causal model grows with the design stage detailing and allows for incomplete input for inference of answering “what-if” questions.}
	\label{fig:Picture5.jpg}
\end{figure}

Based on the advantages mentioned above, we propose a four-step framework consisting of causal relationships learning and causal effect learning as a realization of causal inference in the building design process. In this framework, we use the combination of \textit{graphical representation in SCMs to intuitively process the causal relationships}, and use the \textit{potential outcome framework (Robin causal model) to formalize quantitative effects}, as presented in Figure \ref{fig:Picture5.jpg}. Especially, the extraction of the causal DAG enables the interconnection with the experts’ domain knowledge. Especially, the pruning and modification with domain knowledge is an important step in relating data-driven information. In this way, the causal DAG provides a reusable bridge to connect and augment design and engineering reasoning. 

For an intuitive understanding, we put this framework into a process scenario description: a construction project starts by identifying boundary conditions, and focusing on deciding upon the building’s footprint, construction type, and shape. At this phase, the available information mainly describes the outer shell with various assumptions (associated with uncertainty) corresponding to the design intention or, in some cases, reversal alternative design. The objective of such assumptions normally is to have a lower cost/energy consumption/environmental impact, which in practice, the precise and informed scenarios via causal models could balance with the cost of making changes and guide further decision-making (sub-graph c, Figure \ref{fig:Picture4.jpg}). The extracted causal knowledge lays down a sufficient structure that can be enriched by design assumptions and intentions. As the design progresses, additional information becomes available, and the uncertainties decrease. The outcome: DAGs store and represent causal principles and outline the boundary of the current design phase; With the ongoing design process, DAGs are expanded or modified to adapt new design phase for further assistance analysis (Figure \ref{fig:Picture5.jpg}). 

In summary, the combined sections \ref{Sec2}\&\ref{Sec3} introduces the causal inference into the design assistance by four central pillars:
\begin{enumerate}
\item An analogy between personal experience and physical knowledge provides a channel for integrating data-driven and knowledge-based methods through causal DAGs. We aim to make designers focus on empirical exploration, reducing the need for repetitive inputs of modeling causal physical knowledge. This separation would achieve a fast cross-sectional examination and avoid conducting erroneous conclusions.
\item Causal model provides a data-driven knowledge extraction method for design process analysis with reduced computational difficulty; Furthermore, we clarify different design assistance dimensions. The causal model allows users to quickly check potential design alternatives in a higher dimension.
\item We clarify the boundary of design assistance based on DAGs. The growth of DAG with reduced uncertainties aligns with the nature of the design process.
\item A four-step framework is proposed to implement causal inference into the design domain with causal structure finding and causal relationship quantification.
\end{enumerate}

\section{Framework implementation: illustrative example}\label{Sec4}

To implement the framework mentioned above, we consider an intuitive energy performance evaluation case in the building design phase. This case focuses on specific geometric and semantic features for a comprehensive understanding. Design decisions have a substantial impact on the resultant energy performance in the early phases, and at the same time hard to evaluate alternatives in the detailed design phase as it might require reconstructing simulation fundamentally. 

In this context, design-sequential embedded causal inference owns advantages and flexibility to make adjustment assumptions.

\subsection{Data}

Illustrating the counterfactual/assumption causal case with validation is a widespread difficulty in causal inference research. The one advantage of our domain in adopting the causal model research is well-developed, sophisticated first-principles simulations, which encode rule-based knowledge/causality through efforts by engineers and experts to provide a solid foundation for data-driven research and counterfactual validation.

In this case study, we illustrate how the causal inference and model extract knowledge from the simulation data and benefit decision-making support. The dataset contains 1000 simulations in various scenarios that were targeted in the climate zone of Munich, Germany \citep{singh2020quick}, generated via EnergyPlus (EP) \citep{crawley2000energy}. We chose this dataset for the case study based on two major reasons:

\begin{enumerate}
\item The dataset emphasizes buildings in different geometric properties. It serves for energy performance evaluation in the early design phase, which contains relatively simple building parametric representations with comprehensive causal relationships that could be examined by our common sense. The first-principles simulation process itself provides verification for extracted knowledge. 
\item The accuracy of performance results from this dataset is tested through a real-world case study \citep{geyer2021explainable}, which means that it is solid for validating the causal effect estimation. 
\end{enumerate}

The selected parameters from the dataset with their corresponding sampling ranges are presented in Table \ref{tab:table1}.

\hspace*{-1.5cm}
\begin{table}[t]
    \normalsize
    \centering
    \begin{tabular}{llll} 
    \hline
    Parameter                             & Unit                                       & Min  & Max\\ 
    \hline
    Ground Floor Area \textsuperscript{1} & m\textsuperscript{2}                       & 250  & 800                                                                                                                          \\ 
    \hline
    Height                                & m                                          & 3    & 4                                                                                                                            \\ 
    \hline
    Number of
      Floors                    & -                                          & 2    & 5                                                                                                                            \\ 
    \hline
    External Wall Area                    & m\textsuperscript{2}                       & 200  & 1800                                                                                                                         \\ 
    \hline
    u-Value
      (Wall)                      & {W/m\textsuperscript{2}K}   & 0.15 & 0.25                                                                                                                         \\
    u-Value
      (Internal Wall)             &                                            & 0.4  & 0.6                                                                                                                          \\
    u-Value
      (Ground Floor)              &                                            & 0.15 & 0.25                                                                                                                         \\
    u-Value
      (Roof)                      &                                            & 0.15 & 0.25                                                                                                                         \\
    u-Value
      (Internal Floor)            &                                            & 0.4  & 0.6                                                                                                                          \\
    u-Value
      (Windows)                   &                                            & 0.7  & 1.0                                                                                                                          \\ 
    \hline
    g-Value
      (Windows)                   & -                                          & 0.3  & 0.6                                                                                                                          \\ 
    \hline
    Permeability                          & m\textsuperscript{3}/m\textsuperscript{2}h & 6    & 9                                                                                                                            \\ 
    \hline
    WWR \textsuperscript{2}               & -                                          & 0.1  & 0.5                                                                                                                          \\ 
    \hline
    Light Heat
      Gain                     & {W/m\textsuperscript{2}}    & 6    & 10                                                                                                                           \\
    Equipment
      Heat Gain                 &                                            & 10   & 14                                                                                                                           \\ 
    \hline
    Building
      Occupancy                  & Person/m\textsuperscript{2}                & 16   & 24                                                                                                                           \\ 
    \hline
    \multicolumn{4}{l}{\begin{tabular}[c]{@{}l@{}}\textsuperscript{1} Ground Floor Area for random shapes buildings \\\textsuperscript{2} Window-to-wall ratio (WWR) varies independently in each direction\end{tabular}}   
    \end{tabular}
    \caption{Parameters and their ranges for the generated dataset, representing the early design phase.}
    \label{tab:table1}
\end{table}

\subsection{Causal-informed assistance, part 1: causal skeleton determination }

The causal inference starts with the causal structure finding. In this subsection, we use the case to illustrate steps 1 and 2 from Figure \ref{fig:Picture5.jpg} framework: Discover skeleton by causal relationship finding algorithm, and how domain-knowledge helps to prune relationships. 

\begin{figure}[h]
	\centering
	\includegraphics[width=\textwidth]{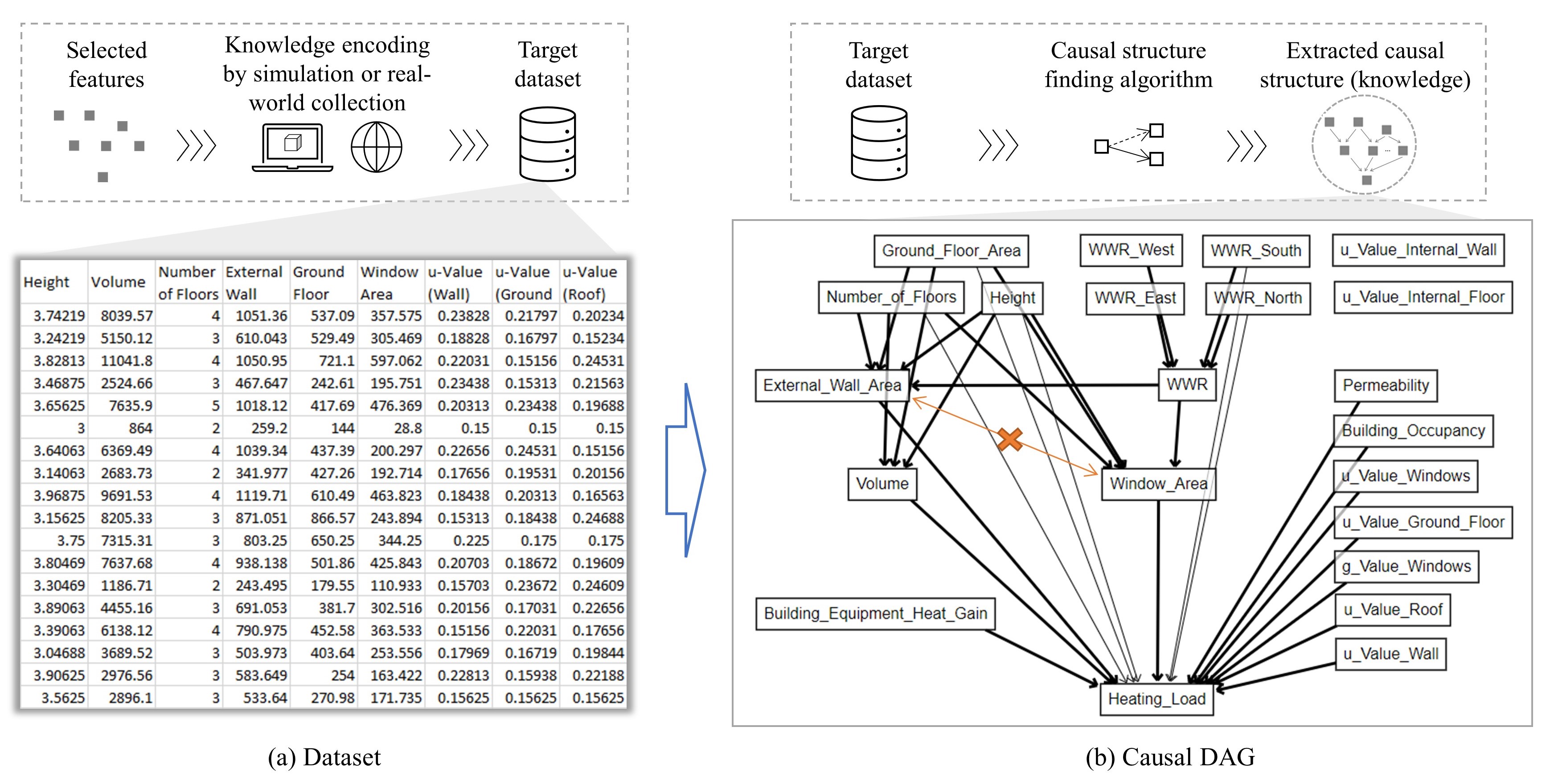}
	\caption{Causal structure finding via GES: knowledge extraction from data in the building’s early design phase. (a) The data generation process: Via knowledge-based simulations or real-world data collection, causal relationships have been implicitly encoded in the dataset. In this figure, the presented dataset is generated by Table 1 configurations with real-world case validation \citep{geyer2021explainable,singh2020quick}. (b)The mechanism of the causal structure finding algorithm is designed to extract casual relationships directly from data and can be further transferred into DAGs. In the skeleton, arrows in bold represent the atomic direct effects, which means removing one of these arrows from the diagram implies that there will no longer be any causal effect between the corresponding variables. }
	\label{fig:Picture6.jpg}
\end{figure}

As mentioned in Section \ref{Sec2}, the GES algorithm is applied to extract causal relationships from the dataset and then presented via DAGitty. An important milestone is that: instead of manually hardcoding rules, a data-driven process is applied to find causal relationships from the simulation dataset, which means we obtain the building physics causality (cause-effect in parameters) in a machine understandable representation. These causal relationships can be extracted directly from data (simulation or real-world collection), which opens opportunities for further augmentative encoding (we can always gather more data) and wider adaptation. The illustrative process is shown in Figure \ref{fig:Picture6.jpg} with relationships in Table \ref{tab:table2}.

The sub-graph (b) of Figure \ref{fig:Picture6.jpg} presents the pruned causal structure result for the dataset generated by Table \ref{tab:table1} configurations (section 4.1) evaluating the building heating load in the DAG skeleton. We need to point out that the extracted skeleton by the algorithm has represented the causal relationships quite accurately and comprehensively. Only slight adjustments are required from the original skeleton by removing the bidirectional arrow between \textit{external wall area} and \textit{window area}. This means that the algorithm fails to conduct the cause-effect direction between these two features from the given data. By applying design domain knowledge, i.e., usually, we design first the building body leading to the external wall; then we design the façade with the fixed window-to-wall ratio (WWR) determining the window area. It does not mean that the external wall area always has a direct correlation to the window area (confounder: WWR). In fact, the need for removing the bidirectional arrow perfectly presents an example of “correlation does not imply causation” in the building design domain.

\begin{table}
\centering
\arrayrulecolor[rgb]{0.498,0.498,0.498}
\begin{tabular}{ll} 
\arrayrulecolor{black}\hline
\textbf{Node (Cause)}                    & \textbf{Descendant (Effect)}                                                  \\ 
\arrayrulecolor[rgb]{0.498,0.498,0.498}\hline
\textit{Height}                          & \multirow{3}{*}{Volume, Heating\_Load, External\_Wall\_Area,
  Window\_Area}  \\ 
\cline{1-1}
\textit{Number\_of\_Floors}              &                                                                               \\ 
\arrayrulecolor{black}\cline{1-1}
\textit{Ground\_Floor\_Area}             &                                                                               \\ 
\arrayrulecolor[rgb]{0.498,0.498,0.498}\hline
\textit{WWR}                             & External\_Wall\_Area, Window\_Area                                            \\ 
\arrayrulecolor{black}\cline{1-1}\arrayrulecolor[rgb]{0.498,0.498,0.498}\cline{2-2}
\textit{WWR\_West}                       & \multirow{2}{*}{WWR}                                                          \\ 
\cline{1-1}
\textit{WWR\_East}                       &                                                                               \\ 
\arrayrulecolor{black}\cline{1-1}\arrayrulecolor[rgb]{0.498,0.498,0.498}\cline{2-2}
\textit{WWR\_South}                      & \multirow{2}{*}{WWR, Heating\_Load}                                           \\ 
\cline{1-1}
\textit{WWR\_North}                      &                                                                               \\ 
\arrayrulecolor{black}\hline
\textit{Volume}                          & \multirow{11}{*}{Heating\_Load}                                               \\ 
\arrayrulecolor[rgb]{0.498,0.498,0.498}\cline{1-1}
\textit{External\_Wall\_Area}            &                                                                               \\ 
\arrayrulecolor{black}\cline{1-1}
\textit{Window\_Area}                    &                                                                               \\ 
\arrayrulecolor[rgb]{0.498,0.498,0.498}\cline{1-1}
\textit{u\_Value\_Wall}                  &                                                                               \\ 
\arrayrulecolor{black}\cline{1-1}
\textit{u\_Value\_Ground\_Floor}         &                                                                               \\ 
\arrayrulecolor[rgb]{0.498,0.498,0.498}\cline{1-1}
\textit{u\_Value\_Roof}                  &                                                                               \\ 
\arrayrulecolor{black}\cline{1-1}
\textit{u\_Value\_Windows}               &                                                                               \\ 
\arrayrulecolor[rgb]{0.498,0.498,0.498}\cline{1-1}
\textit{g\_Value\_Windows}               &                                                                               \\ 
\arrayrulecolor{black}\cline{1-1}
\textit{Permeability}                    &                                                                               \\ 
\arrayrulecolor[rgb]{0.498,0.498,0.498}\cline{1-1}
\textit{Building\_Equipment\_Heat\_Gain} &                                                                               \\ 
\arrayrulecolor{black}\cline{1-1}
\textit{Building\_Occupancy}             &                                                                               \\ 
\arrayrulecolor[rgb]{0.498,0.498,0.498}\hline
\textit{u\_Value\_Internal\_Floor}       & \multirow{2}{*}{None}                                                         \\ 
\arrayrulecolor{black}\cline{1-1}
\textit{u\_Value\_Internal\_Wall}        &                                                                               \\
\arrayrulecolor[rgb]{0.498,0.498,0.498}\hline
\end{tabular}
\arrayrulecolor{black}
\caption{Causal kinships extracted by GES. Most of them can be easily validated by domain knowledge; The process is purely data-driven without rule-based predefinition.}
\label{tab:table2}
\end{table}

In more detail, the following conclusions were observed: 
\begin{enumerate}
\item \textbf{Causal dependence and interpretability}: Most of the causal relationships are correctly presented in DAG, i.e., building geometric features: \textit{height}, \textit{ground floor area}, and the \textit{number of floors} determine the \textit{external wall area} and building \textit{volume}; \textit{WWRs in different directions $\rightarrow$ WWR} combined with building geometric determine the \textit{window area}; Properties features (u-values, g-value, etc.) are independently related to \textit{heating load}, excepts \textit{u-values of internal wall \& floor}.
\item \textbf{Kinships and design process}: The causal skeleton depicts a set of hypotheses about the causal process and offers users the minimal sufficient adjustment sets for estimating the direct effect from treatment to the outcome as a map. These kinship relations reflect the sequence of design process decision-making: \textit{building body with external dimensions $\rightarrow$ façade with WWR $\rightarrow$ property \& material features}. 
\item \textbf{Domain-knowledge integration}: Except for the causal relationship, there are bidirectional correlations that exist in the DAG, i.e. \textit{building external wall area} and \textit{window area}: They are related but not causal because both features are determined by the geometric dimensions and WWRs. In this context, skeleton modification and enrichment via domain knowledge are necessary.
\end{enumerate}

\subsection{Causal-informed assistance, part 2: causal relationships identification}

In this stage, the skeleton encoded causal information provides logistical insights, which helps to examine the complete input independence or their prerequisites (Figure \ref{fig:Picture3.jpg}), and further answer hypothetical questions. In this section, on a case base, we provide the necessary information to address typical assumptions during the design stage based on the previous section’s outcome. We use exemplary questions occurring frequently in building design for that a general answer is not available but individual answers are required. These questions demonstrate the causal inference:
\begin{itemize}
\item \textit{Scenario i}: What if I change the \textit{window area}, how does it change the \textit{heating load}?
\item \textit{Scenario ii}: What if I change the \textit{building floor height}, how does it change the \textit{heating load}?
\end{itemize}

Essentially, DAGs provide valuable information to designers on whether some specific features should be controlled or not, based on the given assumption scenario (target treatment) and the target outcome. The design process knowledge is transformed into the skeleton; this encoding process makes machines understand domain causalities for assistance. Notably, causal rules in DAGs enable machines to correct errors that we could normally neglect in sensitivity analysis. 

\begin{figure}[h]
	\centering
	\includegraphics[width=\textwidth]{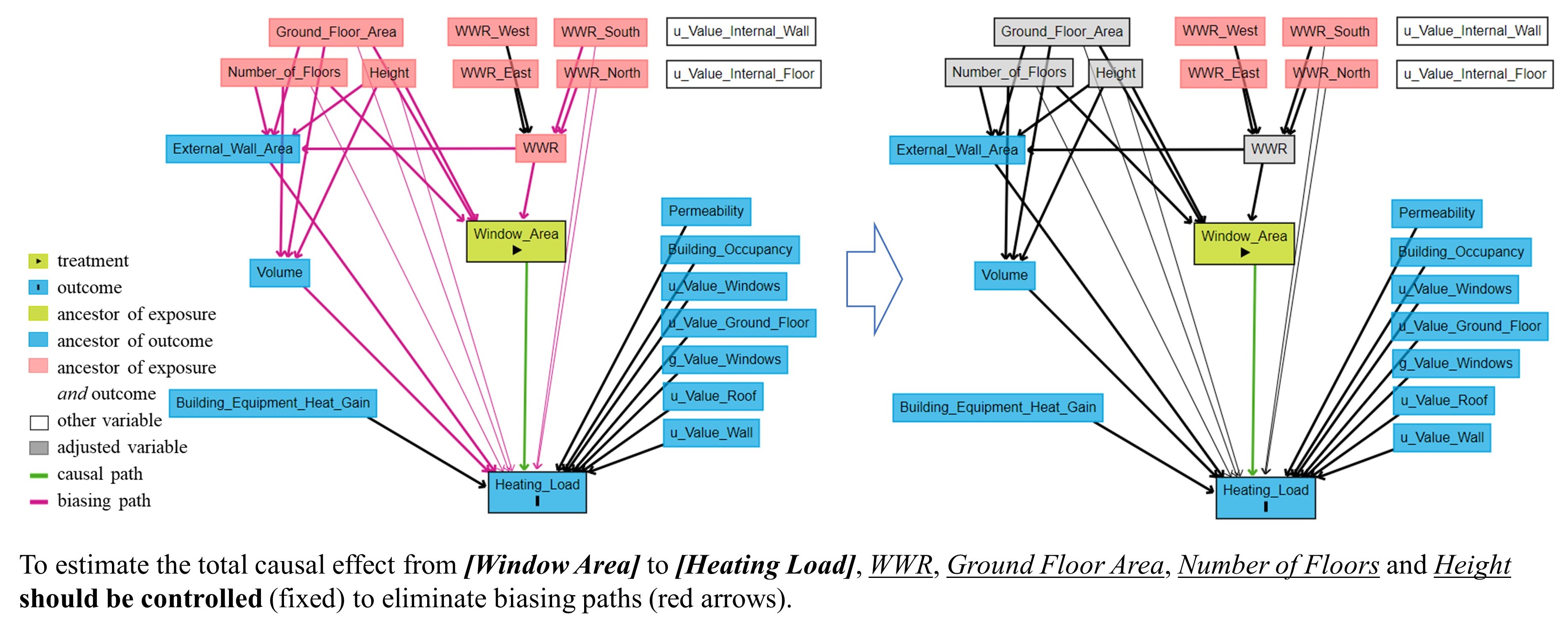}
	\caption{Design process support, Scenario i: If the designer wants to know how the window area would affect the heating load, Rules of causal DAGs suggest that they need to consider controlling the WWR and building floor height, floor area, and the number of floors to get an unbiased effect estimation.}
	\label{fig:Picture7.jpg}
\end{figure}

To address \textit{Scenario i} dealing with window area, we firstly set the treatment variable (\textit{Window Area}) and outcome (\textit{Heating Load}), the skeleton based on the relationships and casual rules to examine whether there are \textbf{biasing paths}. The biasing path means one needs to be blocked by controlling variable(s) in the path to avoid biased results \citep{textor2012adjustment,textor2015drawing}, which we refer to in subfigure (a), Figure \ref{fig:Picture2.jpg}. A process visualization is presented in Figure \ref{fig:Picture7.jpg}.

The rules extracted from DAGs are interpreted as the suggestion:
\begin{itemize}
\item \textit{Suggestion to Scenario i}: To correctly estimate the direct causal effect between \textit{Window Area and Heating Load, Ground Floor Area, Floor Height, Number of Floor}, and \textit{WWR} should be adjusted.
\end{itemize}
An interpretation by domain knowledge is: The causal skeleton only refers to buildings with different facade areas (shapes) but the same volume and WWR to get rid of: 
\begin{enumerate}
\item Confounder from building \textit{Volume}. 
\item Correlation between \textit{External Wall Area} and \textit{Window Area}. 
\end{enumerate}
From the perspective of the design process, the answer delivers a different layer of information: we should only start considering the window area of the building when we have set the abovementioned features.

\begin{figure}[h]
	\centering
	\includegraphics[width=\textwidth]{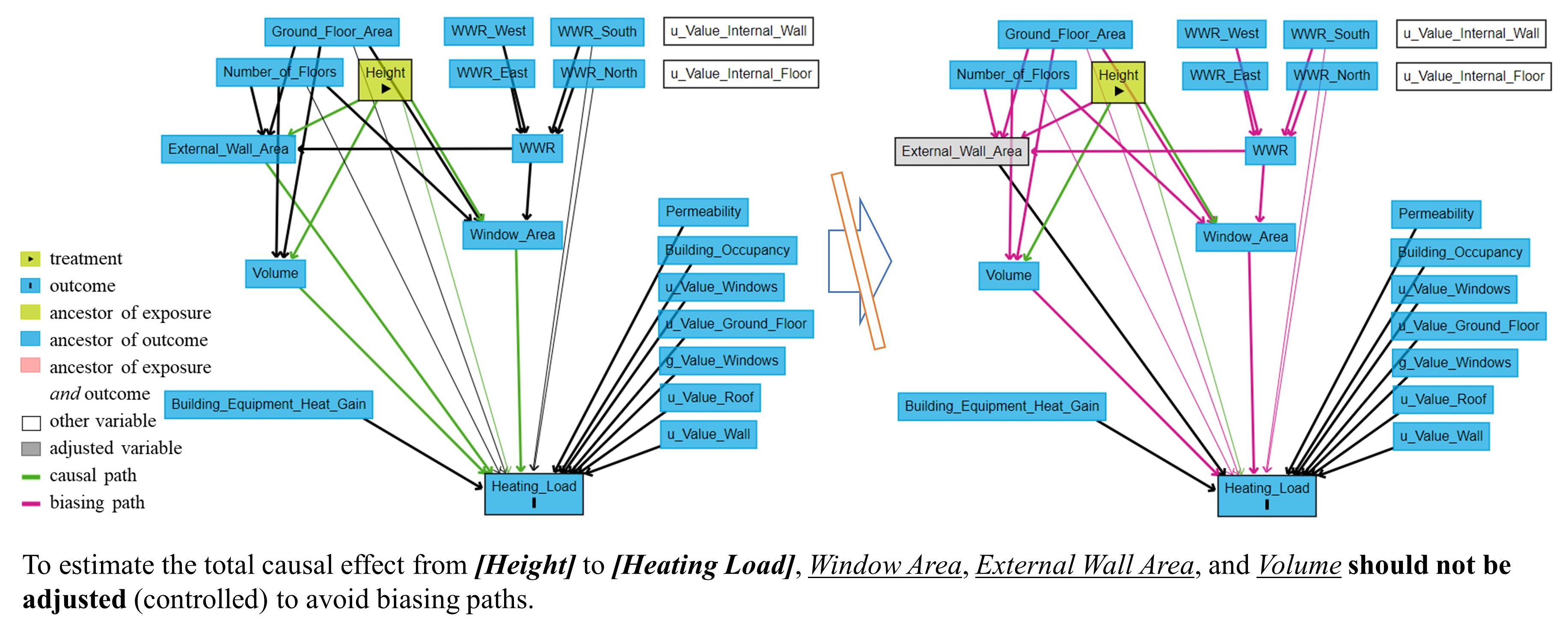}
	\caption{Design process support, Scenario ii: If the designer wants to know how the total effect of height (treatment of exposure) on heating load (outcome), Rules of causal DAGs suggest that they should leave all causal paths open and not cause new biasing paths by controlling features of window area, volume, and external wall area to get unbiased effect estimation.}
	\label{fig:Picture8.jpg}
\end{figure}

Sequentially, Figure \ref{fig:Picture8.jpg} illustrates the analysis process regarding \textit{Scenario ii}.

To address the \textit{Scenario ii}: Investigating how the floor \textit{Height} of a building (treatment) would affect the \textit{Heating Load} (outcome). In this case, the treatment variable, building floor height, owns causal descendants but no ancestors. The important reminder is that we should not infer nodes along direct causal paths (bold arrows in Figure \ref{fig:Picture8.jpg}) from Height to Building Load, to create other bias paths (please refer to the collider situations in the subfigure (b), Figure \ref{fig:Picture2.jpg}). They contain causal relationships with other features, blocking them would involve new biasing paths to the model by controlling them. Hence, the suggestion is as follows:
\begin{itemize}
\item \textit{Suggestion to Scenario ii}: To correctly estimate the direct causal effect between floor \textit{Height} and \textit{Heating Load, Volume, External Wall Area}, and \textit{Window Area} shouldn’t be adjusted.
\end{itemize}

It means that when we consider the direct effect of changing building floor height, the building volume, façade, and especially the window area should change accordingly, to avoid bringing new bias (caused by WWR, floor area, etc.) into the outcome. Naturally, this suggestion is also adaptable to the support of the design process sequence.

Here, we’d to strengthen that the process mentioned above runs in causal modeling  automatically, which is embedded in the process once the skeleton is fixed. The target identification, including treatment, outcome, and conditions extracted based on the causal skeleton via graph-based criteria, is called estimand identification. We refer to this process as Step 3, and the follow-up quantified mathematical formulation and calculation as Step 4 in the four steps of the framework, Figure \ref{fig:Picture5.jpg}. 

\subsection{Causal-informed assistance, part 3: causal effects quantification}

In this subsection, we further investigate and quantify the causal effect by using the potential outcome framework to explain a quantitative scenario during the design process:
\begin{itemize}
\item \textit{“What if I had changed floor height from 3 meters to 3.2 meters, how does it make difference on heating load?”}
\end{itemize}

Once the causal relationship has been clarified between the treatment and the outcome, the investigation of causal effect estimation is relatively intuitive. In this case, we set an experiment based on Figure \ref{fig:Picture8.jpg} and are particularly interested in the putative effect of how the building heating load is influenced by altering floor height via intervention (Step 3, Figure \ref{fig:Picture5.jpg}). To simulate the scenario, we set values to random features and keep others unknown to represent a certain stage of the ongoing design phase, as shown in Table \ref{tab:table3}.

\begin{table}[h]
    \centering
    \normalsize
    \begin{tabular}{lll} 
    \hline
    Parameter                            & Unit                                       & Value    \\ 
    \hline
    Ground Floor Area \textsuperscript{} & m\textsuperscript{2}                       & 300      \\ 
    \hline
    Height                               & m                                          & 3 $\rightarrow$ 3.2  \\ 
    \hline
    Number of
      Floors                   & -                                          & 3        \\ 
    \hline
    External Wall Area                   & m\textsuperscript{2}                       & Unknown  \\ 
    \hline
    u-Value
      (Wall)                     & {W/m\textsuperscript{2}K}   & Unknown  \\
    u-Value
      (Internal Wall)            &                                            & Unknown  \\
    u-Value
      (Ground Floor)             &                                            & 0.2      \\
    u-Value
      (Roof)                     &                                            & 0.2      \\
    u-Value
      (Internal Floor)           &                                            & Unknown  \\
    u-Value
      (Windows)                  &                                            & Unknown  \\ 
    \hline
    g-Value
      (Windows)                  & -                                          & Unknown  \\ 
    \hline
    Permeability                         & m\textsuperscript{3}/m\textsuperscript{2}h & 7.5      \\ 
    \hline
    WWR                                  & -                                          & 0.3      \\ 
    \hline
    Equipment
      Heat Gain                & W/m\textsuperscript{2}                     & Unknown  \\ 
    \hline
    Building
      Occupancy                 & Person/m\textsuperscript{2}                & Unknown  \\
    \hline
    \end{tabular}
    \caption{Causal effect estimation scenario: "what if
I had changed floor height from 3 meters to 3.2 meters?".}
    \label{tab:table3}
    \vspace{0cm}
\end{table}

The counterfactual experiment is designed in the following process: We firstly set building \textit{height} as treatment and \textit{heating load} as the outcome. Based on the result we conclude from the DAGs (section 4.3), the \textit{Volume}, \textit{External Wall Area}, and \textit{Window Area} are in the direct causal path and should remain open. The potential outcome expresses as: 
\[\tau =E[Heating Load\mid Height = 3.2m, \textit{X}]- E[Heating Load\mid Height = 3m, \textit{X}]\]
in which:
\begin{itemize}
\item 	If we calculate the ATE, the \textbf{\textit{X}} should be the set of \textit{[Ground Floor Area, Number of Floors, Building Equipment Heat Gain, Building Occupancy, WWRs, u Value Roof, Ground Floor u Value, Permeability]} sampled in their ranges, independently.
\item 	If we calculate the CATE, the \textbf{\textit{X}} is then adapted with the Table \ref{tab:table3} conditions and becomes the set of \textit{[Building Equipment Heat Gain, Building Occupancy, Ground Floor Area=300, Number of Floors=3, WWRs=0.3, u Value Roof=0.2, u Value Ground Floor=0.2, Permeability=7.5]}
\end{itemize}

For the result evaluation, we use EP to run simulations and validate the accuracy of outcomes by generating ATE dataset and CATE dataset. Both datasets have two different data batches: control height of 3 meters and 3.2 meters. We generated 200 samples for each batch based on set ranges (ATE: ranges from Table \ref{tab:table1}; CATE conditions from Table \ref{tab:table3}) as the ground-truth test set. 

In comparison, we also applied the data-driven method without a causal-informed scenario: We selected three typical ML/AI regression methods widely used in the BPS domain \citep{seyedzadeh2018machine}: Gradient Boost (LightGBM) \citep{ke2017lightgbm}, Random Forest (RF), and Artificial Neural Network (ANN), evaluated by two common metrics: Mean Absolute Percentage Error (MAPE) and R2 coefficient of determination. Table \ref{tab:table4} presents their 4-fold cross-validation \citep{refaeilzadeh2009cross} accuracy scores trained by the dataset described in Table \ref{tab:table1}. 
\begin{table}
\centering
\arrayrulecolor[rgb]{0.498,0.498,0.498}
\begin{tabular}{lll} 
\arrayrulecolor{black}\hline
                  & \textbf{MAPE} & \textbf{R\textsuperscript{2}}  \\ 
\arrayrulecolor[rgb]{0.498,0.498,0.498}\hline
\textbf{LightGBM} & 6.972
  \%    & \textbf{0.924}                          \\ 
\hline
\textbf{RF}       & 11.016 \%     & 0.81                           \\ 
\arrayrulecolor{black}\hline
\textbf{ANN}      & 13.152
  \%   & 0.746                          \\
\arrayrulecolor[rgb]{0.498,0.498,0.498}\hline
\end{tabular}
\arrayrulecolor{black}
\caption{Accuracy performance of three typical data-driven methods trained by the table \ref{tab:table1} dataset for heating load prediction. We used the default setting of models in their open-source code implementation \citep{pedregosa2011scikit,ke2017lightgbm} with a certain range of optimal hyperparameter grid-search conducted. The score is generated by 4-fold cross-validation. LightGBM shows the best performance in prediction. }
\label{tab:table4}
\end{table}

\begin{figure}[h]
	\centering
	\includegraphics[width=\textwidth]{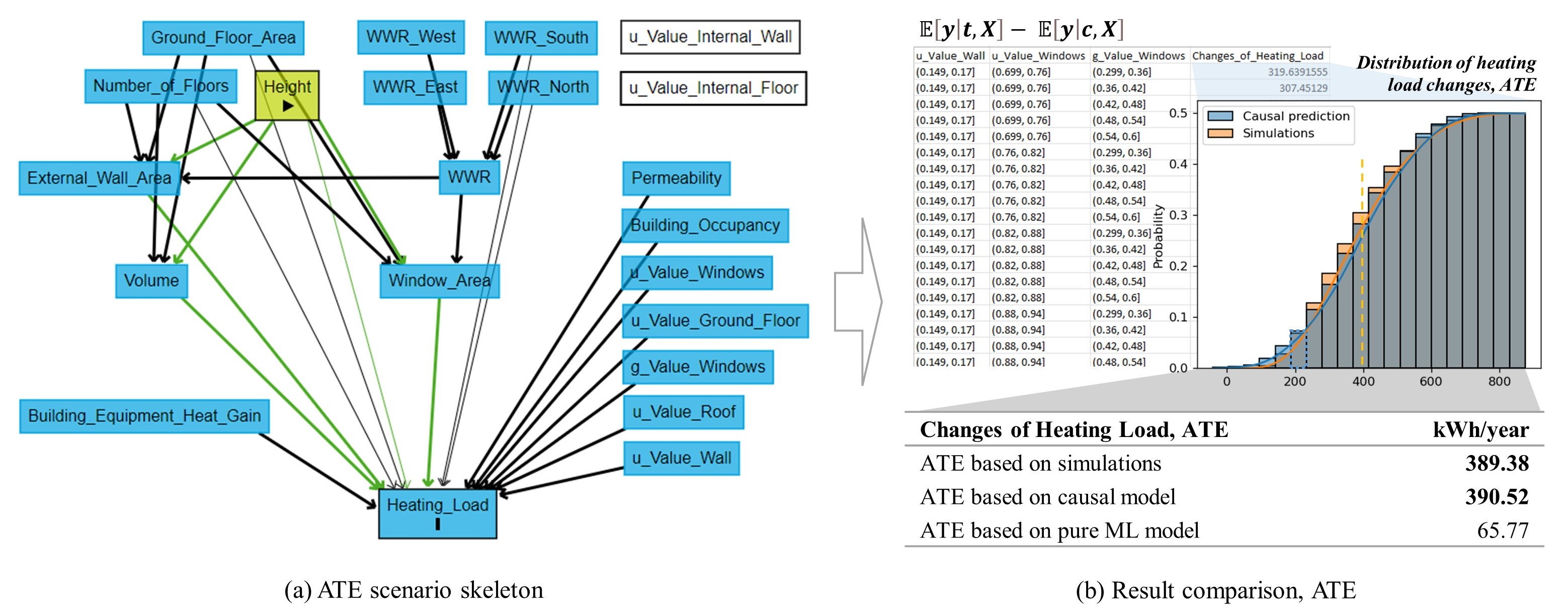}
	\caption{Causal effect estimation process, ATE scenario: (a) Causal skeleton based on Figure \ref{fig:Picture8.jpg} scenario information; (b) Based on the given causal skeleton, the potential outcome framework calculates ATE estimation with its accumulative distribution.}
	\label{fig:Picture9.jpg}
\end{figure}

\begin{figure}[h]
	\centering
	\includegraphics[width=\textwidth]{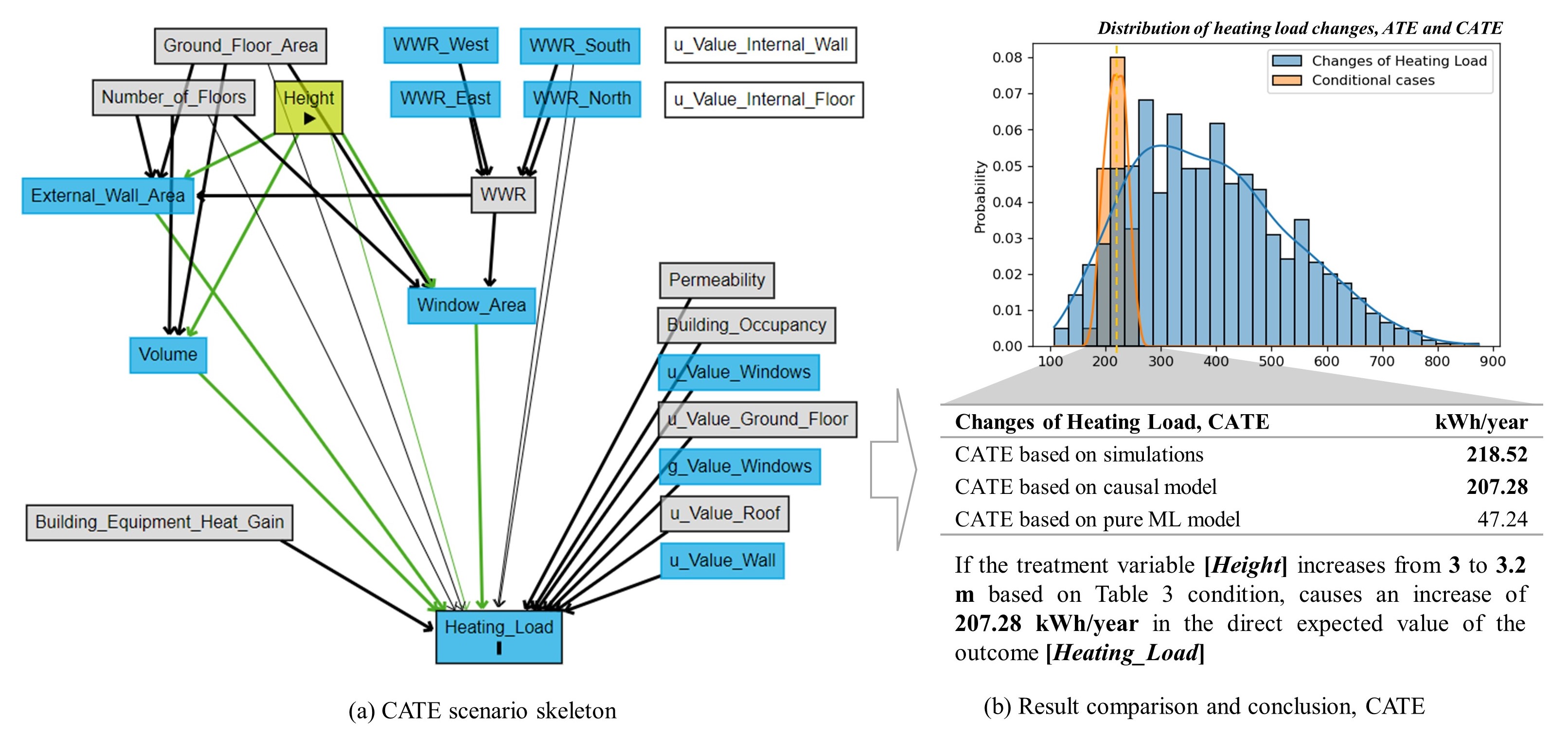}
	\caption{Causal effect estimation process, CATE scenario: (a) Causal skeleton based on Table \ref{tab:table3} condition; (b) From the comparative results resampled from ATE output distribution (Figure \ref{fig:Picture9.jpg}), the causal model presents quantified potential outcomes of CATE and answers the Table \ref{tab:table3} question.}
	\label{fig:Picture10.jpg}
\end{figure}

Based on the performance result, we select LightGBM to run the conditioned data on Table \ref{tab:table2} (ATE) and Table \ref{tab:table3} (CATE) scenarios. To simulate the data-driven process with neglect of feature dependencies, the unknown features are sampled individually (i.i.d.) as their original distribution ranges shown in Table \ref{tab:table1} without considering causal relationships. The outcome comparison between ground-truth simulation, potential outcome framework (causal model), and LightGBM without causal-informed (pure data-driven model) is shown in tables of Figure \ref{fig:Picture9.jpg} and Figure \ref{fig:Picture10.jpg} for both scenarios.

Figure \ref{fig:Picture9.jpg} illustrates the causal estimation process of ATE: The causal model based on the Figure \ref{fig:Picture8.jpg} skeleton conducts the intervention result of the heating load in different conditions. The estimation result conveys critical information for design decision-aids: Within the range of Table \ref{tab:table1}, if we intervene the building height from 3 to 3.2 meters, the heating load changes would vary between 0 - 900 kWh/year. The average changes and the distribution are validated by simulations, which show only minor differences; Based on validated ATE, Figure \ref{fig:Picture10.jpg} presents the condition dependencies from the set feature conditions in Table \ref{tab:table3} and the resampling process from ATE. The CATE shows that the changes should be 207.28 kWh/year on average, with the validated simulation result of 218.52 kWh/year. In both scenarios, the pure ML method without causal-informed fails to produce correct potential outcome ranges to answer “what-if” questions, even though they perform well in the prediction task. The reason is: In the pure ML forecasting process, the input range sampling ignores the constraints of feature dependencies, which leads to unrealistic input combinations involved in the forecasting, causing biased results and wrong information.

To sum up, by encoding causality knowledge into the skeleton with potential outcome estimation, the causal model achieves compact and accurate decision-aid information: The case presents the scenario in an energy-efficient design process, including counterfactual design variants. It is a computationally efficient way than simulations and compensates for the deficiencies of current widely used, data-driven methods.

\section{Discussion}\label{Sec5}

In this section, framework adaptability, limitations, and the prospects of causal inference in the design process are discussed. 

For the framework adaptability, this study uses DAGs, causal discovery algorithms, and simulation to introduce causal inference in the scope of energy-efficient building design. The framework fits the general data-driven process assistance.  Causal inference offers a flexible but mathematical rigor approach to intervene or examine counterfactuals to discover facts in predictive scenarios and adapt strategies \citep{cao2016building}; in other words, it serves as Explainable Artificial intelligence (XAI) \citep{li2020survey} in a causal-aware manner for model interpretation. We intend to raise a call for the energy domain and building design community to pay more attention to reviewing ML methodologies and examining the parametric independence for decision-making support. 

For the limitation, obstacles to causal inference application in our domain come from two major aspects: 1. The inherent data-driven nature of the method, and 2. The validation of counterfactuals. Causal models are essentially data-hungry as a complement to statistical methods. In this context, the integration of causal models in our domain is advantageous due to the well-fundamental first-principles models and simulation tools developed by prior knowledge. In this study, we use a dataset from the BPS description to extract knowledge based on physical causal relations to conduct the design process; however, to our best knowledge, a gap regarding the mutual parameter adaptability still exists between tools in BIM and BPS communities for building and process representation, which causes difficulties in data acquisition from the design process in BIM or mutual transfer for design process integration. Further effort in the methods regarding data acquisition or enrichment to describe the design process, design parameters, and causal relationships are meaningful. Particularly, the dataset for the test of counterfactual, and the combination of simulated data in variance with real-world data validation are required for further evaluation support. 

If we consider design cases in reality, absurd conclusions from Figure \ref{fig:Picture2.jpg} are less likely to occur in individual design cases when there are only a few variables. Architects and engineers could generally rule out absurd conclusions by their domain knowledge or experience and make further decisions. Those decisions in the early design phase will significantly shape the building performance, which during the design process will derivative many “what-if” questions for particular alternative feature importance analysis; however, when the design case is complex (multiple confounder and collider combined), or when the building designer lack of relevant domain knowledge, oversights or wrong decisions are highly possible to occur by data-driven decision-making support tools without investigation of causal relationships.

In our case study, we only investigated causal relationships by involving basic building characteristics and statistics energy performance. If we consider the scope of the energy-efficient building design, further research is worth conducting to gain causal insights by involving more factors: external conditions (e.g., weather, geography), life-cycle assessment, internal influences (e.g., user behaviors), causality analysis in dynamic time-series data, or even perspectives from design cognition domain, etc. In the community, we see successful symbolic-based approaches such as energy-emergy integration for building shape optimization \citep{yi2015integrated} by unifying different objectives into the energetic flow form to support design decision-making. It could be a knowledge-based way to reconcile different parametric representations into the same context for data enrichment. However, from another perspective, the potential risk of wrong conclusions by applying current analysis tools is increasing: the development of simulations and the spread of digitalization are raising the data volume dramatically with the trend of interdisciplinary requirements. In this context, we argue that causal inference is not only beneficial but necessary to be involved in any design assistance; The knowledge discovery via DAG structure finding from data would effectively help architects to make better decisions and avoid wrong judgments. Since the knowledge is stackable and transferable in different design cases, establishing general knowledge libraries by collective intelligence contains huge potential, which would help every designer to reach informed optimal scenarios in each “what-if” question during the process, simultaneously.

\section{Conclusion}\label{Sec6}

In this study, we introduce causal inference into the sustainable building design domain for the first time and propose a two-part process to construct the causal model: 1. Using DAG from SCMs to discover causal relationships from the set of design parameters representative of the design scheme; 2. Developing estimators to evaluate a given treatment effect by combing DAGs and potential outcomes framework. The framework design aims to involve causal analysis in the parametric design process. It provides a computationally efficient behavior with extendable structures and mathematical rigor rules for inferring consequences under conditions that changes during the design: induced by treatments or external interventions. Eventually, allow the machine assistance to discover and extract causal knowledge from data directly without semantic grounding required, conduct “what-if” questions, reveal correct design alternatives, and eventually, reach energy-efficient optimal with causal-informed design assistance. At the same time, the idea of differentiating knowledge and experience based on reusability needs to be integrated into the design assistance. DAGs offer an excellent medium for the general knowledge storage purpose for the following reasons:

\begin{itemize}
\item DAGs provide a complete mathematical language for describing and utilizing invariant knowledge, principles, and design sequences in different design schemes. More importantly, the generation and modification of DAGs are supported by causal structure finding algorithms and domain knowledge.
\item To address “what-if” questions during the design process, the structure of the DAG describes a clear boundary for alternative generation within reasonable design space. With set target treatment(s) and output, DAGs solve the design parameter entanglement issues by providing interpretable information on which features need to be conditioned.
\item With the extensibility of the DAG, it is possible to further supplement general knowledge from different design cases and provide the basis for incremental learning or collective intelligence for consensus decision-making. 
\item The methodology opens a possibility to encode and express causal inference in a rigorous mathematical way. As a data-driven process, it owns the huge potential to expand in scale, and adapts different aspects to support sophisticated, advanced design decision-making.
\end{itemize}

Nowadays, the rapid advancements of AI have attracted attention in most fields; however, due to its inherent connectionism structure and data-driven nature, significant progress has been made in areas where the data fundamental is more advantageous, such as images, speech, and text processing. Nevertheless, many domains are still dominated by empirical disciplines and principles, such as architecture. In this context, it is worth exploring how to transfer our knowledge that enables ML algorithms to exploit instead of data pattern induction. The globally shared vision of sustainability raises challenges for designers and engineers to equip with interdisciplinary professions due to the requirement from different aspects: life-cycle assessment, environmental impact, cost, etc. The big picture of future development indicates that the knowledge and information volume will keep increasing. In this context, the causal model provides a new and necessary foundation to reconcile the knowledge in a mathematical rigor and computationally efficient way to combine it with a data-driven process. By extending the graphical-based causal rules and potential outcome estimation, the framework would cover a more comprehensive field and become a necessity. Thereby supporting designers and engineers to design more efficiently and to reach beyond their personal capabilities.

DAGs with causal inference offer a possibility to achieve this integration and remind us to distinguish between correlation and causation. For answering “what-if” questions and toward higher intelligence of machine assistance in the empirical science domain, more exploration, effort, and practice are required for knowledge abstraction and reformatting for deeper compilation. Interestingly, the extraction of causal structures from the data gives a form of interpretability that is very close to the common design reasoning process. In this way, it reveals an intuitive path with a higher degree to guide the data exploration and interpretation process for bridging interdisciplinary professions. The causal model provides a methodology for rules abstraction based on data and an efficient medium of compilation and reformatting knowledge. In this context, the proposed two-part process inherits the advantages of symbolism and plays a complementary role to data-driven approaches. We believe this is one of the potential solutions to extend digitalization and the benefit of artificial intelligence into the empirical science domain. To further explore the domain integration of causal modeling, requires further examination of the nature of compiled representations, intrinsic limitations, the types of reasoning they support, and the effectiveness in getting the answers that users expect to get.

\section{Acknowledgements}
We gratefully acknowledge the German Research Foundation (DFG) support for funding the project under grant GE 1652/3-2 in the Researcher Unit FOR 2363.

\bibliographystyle{unsrtnat}
\bibliography{references} 

\end{document}